\documentclass[aps,prl,twocolumn,superscriptaddress,floatfix]{revtex4}
\usepackage{times,amsfonts,amsmath,amssymb,graphicx,graphics,color}

\begin{document}

\title{Density-Dependent Synthetic Gauge Fields Using Periodically Modulated Interactions}

\author{S. Greschner}
\affiliation{Institut f\"ur Theoretische Physik, Leibniz Universit\"at Hannover, Appelstr. 2, DE-30167 Hannover, Germany}

\author{G. Sun}
\affiliation{Institut f\"ur Theoretische Physik, Leibniz Universit\"at Hannover, Appelstr. 2, DE-30167 Hannover, Germany}

\author{D. Poletti}
\affiliation{Engineering Product Development, Singapore University of Technology and Design, 20 Dover Drive, 138682 Singapore}
\affiliation{Merlion MajuLab, CNRS-UNS-NUS-NTU International Joint Research Unit, UMI 3654, Singapore}

\author{L. Santos}
\affiliation{Institut f\"ur Theoretische Physik, Leibniz Universit\"at Hannover, Appelstr. 2, DE-30167 Hannover, Germany}

\begin{abstract}
We show that density-dependent synthetic gauge fields may be engineered by combining periodically modulated interactions and 
Raman-assisted hopping in spin-dependent optical lattices. These fields lead to a density-dependent shift of the momentum distribution, may induce 
superfluid-to-Mott insulator transitions, and strongly modify correlations in the superfluid regime. 
We show that the interplay between the created gauge field and the broken sublattice symmetry  
results, as well, in an intriguing behavior at vanishing interactions, characterized by the appearance of a fractional Mott insulator. 
\end{abstract}


\maketitle


The emulation of synthetic electromagnetism in cold neutral gases has attracted a major interest~\cite{Dalibard2011,Goldman2013}. 
Artificial electric and magnetic fields have been induced using lasers~\cite{Lin2009,Lin2009b,Lin2011}. Moreover, 
these setups may be extended to generate non-Abelian fields, and in particular spin-orbit coupling~\cite{Lin2011b,Wang2012,Cheuk2012,Zhang2012,Fu2013,Zhang2013,Qu2013,LeBlanc2013}. 
Synthetic fields may be generated as well in optical lattices, and recent experiments have created artificial 
staggered~\cite{Aidelsburger2011, Jimenez-Garcia2012,Struck2012} and uniform~\cite{Aidelsburger2013,Miyake2013} magnetic fields.
These fields are however static, as they are not influenced by the atoms.

The dynamical feedback between matter and gauge fields plays, however, an important role 
in various areas of physics, ranging from condensed-matter~\cite{Levin2005} to quantum chromodynamics~\cite{Kogut1983}, and its realization 
in cold lattice gases is attracting a growing attention~\cite{Wiese2013}.
Schemes have been recently proposed for multi-component lattice gases, such that the low-energy description of these systems is that 
of relevant quantum field theories~\cite{Cirac2010,Zohar2011,Kapit2011,Zohar2012,Banerjee2012,Tagliacozzo2012,Zohar2013,Banerjee2013,Zohar2013b,Tagliacozzo2013}. 
The back-action of the atoms on the value of a synthetic gauge field is expected to lead to interesting physics, including statistically-induced phase transitions and anyons in 1D lattices~\cite{Keilmann2011}, 
and chiral solitons in Bose-Einstein condensates~\cite{Edmonds2013}.

Periodically modulated optical lattices open interesting possibilities for the engineering of lattice 
gases~\cite{Struck2012,Aidelsburger2013,Miyake2013,Eckardt2005,Lignier2007,Kierig2008,Zenesini2009,Struck2011,
Chen2011,Ma2011}.  
In particular, periodic lattice shaking results in a modified hopping rate~\cite{Eckardt2005,Lignier2007,Kierig2008}, which has been employed 
to drive the superfluid~(SF) to Mott insulator~(MI) transition~\cite{Zenesini2009}, to simulate frustrated classical magnetism~\cite{Struck2011}, and to create tunable gauge potentials~\cite{Struck2012}. 
Interestingly, a periodically modulated magnetic field may be employed in the vicinity of a Feshbach resonance to induce 
periodically modulated interactions, which result in a non-linear hopping rate that depends on the occupation differences 
at neighboring sites~\cite{Gong2009,Rapp2012,DiLiberto2013}. 

In this Letter, we show that combining periodic interactions and Raman-assisted hopping may induce a density-dependent gauge field in 
1D lattices. The created field results in a density-dependent shift of the momentum distribution that may be probed 
in time-of-flight~(TOF) experiments. Moreover, contrary to the Peierls phase induced in shaken lattices~\cite{Struck2012}, 
the created field cannot be gauged out, and hence affects significantly the ground-state properties of the lattice gas, leading to 
gauge-induced SF to MI transitions, the emergence of MI at vanishing interaction, and strongly modified correlations in the SF regime.

 \begin{figure}[t]
 \begin{center}
 \includegraphics[width =\columnwidth]{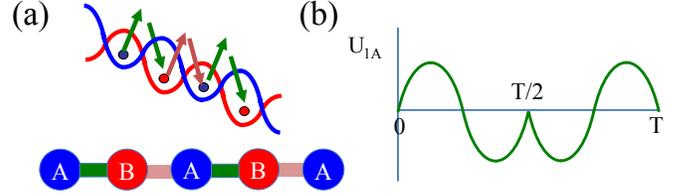}
 \caption{(Color online) Scheme of the AB set up: (a) for $0<t<T/2$ Raman assisted hopping couples an A site with the B site at their right; for $T/2<t<T$ it couples 
 an A site with the B site at their left; (b) the $U_{A1}(t)$ function is $\sin(\omega_{AB} t)$ for $0<t<T/2$ and $-\sin(\omega_{AB} t)$ for $T/2<t<T$, with $\omega_{AB}=4\pi/T$.}
 \label{fig:1}
 \end{center}
 \end{figure}


\emph{AB model.--} We introduce in the following a possible set-up that creates a density-dependent Peierls phase that cannot be gauged out. 
We consider a tilted 1D spin-dependent lattice~(see Fig.~\ref{fig:1}), in which atoms in state $|1\rangle$~$\left (|2\rangle\right )$ are 
confined in the sublattice $A$~($B$). A first pair of Raman lasers induces Raman-assisted hopping between an A site and the B site to 
its right, whereas a second pair leads to hopping between an A site and the B site to its 
left~\cite{footnote-Jaksch}.  We consider that within a period $T$, for $0<t<T/2$ the Raman assisted coupling AB~(BA) is on~(off) and vice versa for $T/2<t<T$. 
The Hamiltonian of the system is:
\begin{eqnarray}
 &&\!\!\!\!\! \hat H^{AB}\!=\!-\!\sum_j \left [ J_{AB}(t) \hat b_{2j}^\dag \hat b_{2j+1} \!+\! J_{BA}(t) \hat b_{2j}^\dag \hat b_{2j-1} \!+\! {\rm h.c.}\right ] \nonumber \\
 && \!\!\!\!\! + \frac{U_A(t)}{2}\sum_j \! \hat n_{2j} (\hat n_{2j}\!-\!1)\!+\!\frac{U_B}{2}\sum_j \! \hat n_{2j+1} (\hat n_{2j+1}\!-\!1).
 \label{eq:HAB(t)}
\end{eqnarray}
where $J_{AB}=J$ and $J_{BA}=0$ for $0<t<T/2$, $J_{AB}=0$ and $J_{BA}=J$ for $T/2<t<T$, and 
even~(odd) site index corresponds to the $A$~($B$) sublattice.
We consider that the interaction of components $|1\rangle$ can be
independently modulated from those of $|2\rangle$, such that 
$U_A=U_{A0}+U_{A1}(t)$, with $U_{A1}(t)=U_{A1}(t+T)$ and $\int_t^{t+T} dt' U_{A1}(t')=0$, 
whereas $U_B$ is constant~(we consider for simplicity $U_{A0}=U_{B}\equiv U$ \cite{footnote-interaction}).  
As shown in Refs.~\cite{Gong2009,Rapp2012} a sufficiently fast modulation of the interactions leads to an effective model with a density-dependent hopping~(as discussed in the Supplemental Material~\cite{SM}, just modulating the interactions in a standard Bose-Hubbard model does result in a density-dependent Peierls phase, but this phase can be gauged out~\cite{SM}) . 
For the particular case of the AB model we obtain for a fast modulation the effective Hamiltonian~\cite{footnote-anyon}: 
\begin{eqnarray}
\!\!\!\!\! \hat H^{AB}_{\rm eff}\!&=&\!-\!\sum_j \left [ \hat b_{2j}^\dag \tilde J_{AB}(\hat n_{2j})\hat b_{2j+1} \!+\!  \hat b_{2j}^\dag \tilde J_{BA}(\hat n_{2j}) \hat b_{2j-1} \!+\! {\rm h.c.}\right ] \nonumber \\
 && \!\!\!\!\! + \frac{U}{2}\sum_j \! \hat n_{2j} (\hat n_{2j}\!-\!1)\!+\!\frac{U}{2}\sum_j \! \hat n_{2j+1} (\hat n_{2j+1}\!-\!1),
 \label{eq:Heff}
\end{eqnarray}
with $\tilde J_{AB}(\hat n_{2j})=\frac{J}{T}\int_0^{T/2} dt\, e^{\mathrm{i}V(t)\hat n_{2j}/\hbar}$,
$\tilde J_{BA}(\hat n_{2j})=\frac{J}{T}\int_0^{T/2} dt\, e^{\mathrm{i}V(t+T/2)\hat n_{2j}/\hbar}$, and $V(t)=\int_0^t U_{A1}(t')dt'$.

For $U_{A1}(t)=\tilde U_{A1}\sin(\omega_{AB} t)$ for $0<t<T/2$~(with $\omega_{AB}=4\pi/T$), and 
$U_{A1}(t)=-\tilde U_{A1}\sin(\omega_{AB}t)$ for $T/2<t<T$~(see Fig.~\ref{fig:1}(b)), 
$\tilde J_{AB}(\hat n_{2j})=\frac{J}{2}J_0(\Omega_{AB}\, \hat n_{2j})e^{\mathrm{i}\Omega_{AB}\,\hat n_{2j}}$, whereas 
$\tilde J_{BA}(\hat n_{2j})=\tilde J_{AB}(\hat n_{2j})^*$, with 
$\Omega_{AB}=\tilde U_{A1}/\hbar\omega_{AB}$. For more general forms of $U_{A1}(t)$~\cite{SM},  
$\arg [\tilde J_{AB}]=\phi_{AB}\hat n_{2j}$ and  $\arg [\tilde J_{BA}]=\phi_{BA}\hat n_{2j}$.
The created Peierls phase cannot be gauged out if $\Phi\equiv\phi_{AB}-\phi_{BA}\neq 0$, crucially altering the ground-state properties.

 \begin{figure} [t]
 \begin{center}
  \includegraphics[width =\columnwidth]{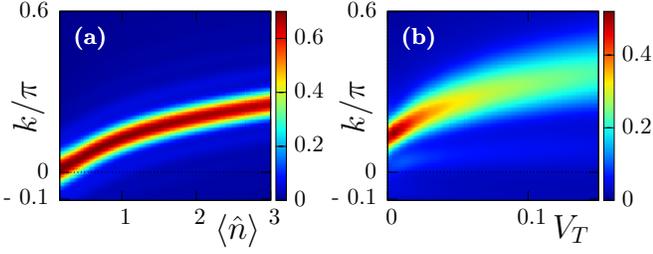} 
 \caption{(Color online) (a) Ground-state quasi-momentum distribution for model~\eqref{eq:Heff} for an homogeneous distribution in $24$ sites with 
 $\Omega_{AB}=\pi/4$, $U=0.2J$, and a density $\langle \hat n \rangle $; (b) same for a harmonically trapped gas as a function of $V_T$~(see text) 
 for $\Omega_{AB}=\pi/4$, $U=J$ and $24$ particles in $24$ sites. Both figures show 
 density-matrix renormalization group~(DMRG)~\cite{Schollwock2011} results with $500$ states, and a maximal occupation per site $n_{max}=10$.}
 \label{fig:2}
 \end{center}
 \end{figure}


\emph{Quasi-momentum distribution.--}  The created Peierls phase results in a drift of the quasi-momentum distribution in the SF regime.
As in recent experiments on shaken lattices~\cite{Struck2012}, this shift may be probed in TOF~(details about experimental detection are discussed below). 
Fig.~\ref{fig:2}(a) shows the quasi-momentum distribution as a function of 
the average density $\langle \hat n \rangle$ for an homogeneous system with $\Omega_{AB}=\pi/4$ and $U=0.2J$. 
However, in contrast to shaken lattice experiments, the momentum shift is density dependent. 
This dependence results in a non-trivial behavior of the quasi-momentum distribution in the presence of 
 an external harmonic confinement, which 
may be accounted for by an additional term $V_T\sum_j (j-L/2)^2 \hat n_j$ in the Hamiltonian~\eqref{eq:Heff}. 
As shown in Fig.~\ref{fig:2}(b),  for larger $V_T$ the quasi-momentum distribution shifts due to growing central density, and broadens due to the inhomogeneous density 
distribution $\langle \hat n_j \rangle$.

 \begin{figure*} [t]
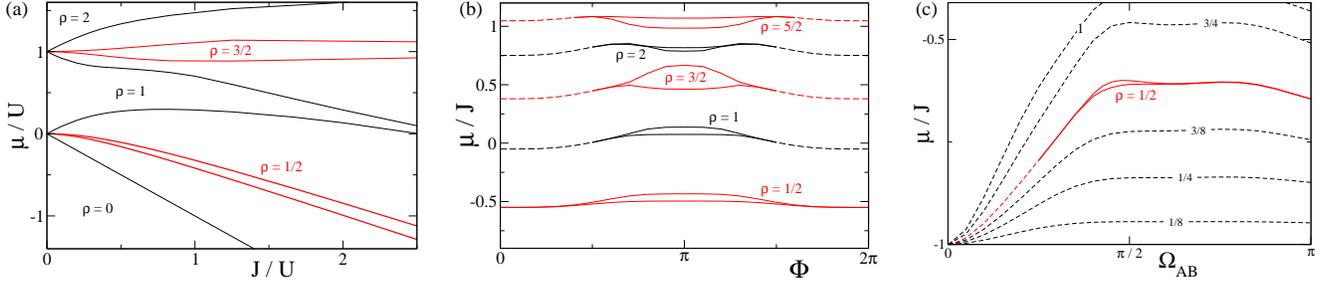

 \begin{center}
\includegraphics[height=3.7cm]{pd_ABBHj_mu_t_q.eps}\hspace{0.4cm}
\includegraphics[height=3.7cm]{lobes_u05_omega157079632679_jphase2_q.eps} \hspace{0.3cm}
\includegraphics[height=3.7cm]{lobes_U0j_dmrg_q.eps} 
 \caption{(Color online) Mott phases at half-integer and integer fillings of model (\ref{eq:Heff}). (a) MI-lobes for $\Omega_{AB}=\pi/2$, $\Phi=\pi$. (b) Varying the relative phase $\Phi$ may induce phase transitions in the ground state. Here we choose $\Omega_{AB}=\pi/2$ and change $\Phi$ \cite{SM} for $J/U=2$ (dashed lines indicate a closing gap) (c) Lines of constant density and MI phase at half-filling for vanishing on-site interactions $U=0$. In the DMRG-calculation system size $L$ and maximal occupation number of bosons per site $n_{max}$ have been scaled carefully (up to $n_{max}=12$ and $L=144$ sites) till a convergence was reached.}
 \label{fig:3}
 \end{center}
 \end{figure*}

\emph{Ground-state phase diagram.--} 
The non-gaugeable density-dependent Peierls phase and the associated broken AB symmetry are crucial for the ground-state physics 
of the AB model~(see Fig.~\ref{fig:3} in which $\mu$ is the chemical potential). MI phases at half-integer filling are induced by the AB asymmetry, 
opening immediately at any finite $J$. For $\langle \hat n \rangle=1/2$ at $J/U\ll 1$ we may project on the 
manifold with $0$ or $1$ particle per site and we may identify $\left|0\right>\to\left|\uparrow\right>$ and $\left|1\right>\to\left|\downarrow\right>$, obtaining up to ${\cal O}(J^2/U)$ the effective spin-$\frac{1}{2}$ Hamiltonian $\hat H_{1/2}=\hat H_0+\hat H_2$, with $\hat H_0=-J\sum_j \hat S^+_j\hat S^-_{j+1}+{\rm h.c.}$, and $(U/J^2) \hat H_2=\sum_j [ \hat S^+_{2j} \left(\frac{1}{2} + \hat S^z_{2j+1} \right) \hat S^-_{2j+2} + \Gamma^2 \hat S^+_{2j-1} \left(\frac{1}{2} + \hat S^z_{2j} \right)\hat S^-_{2j+1}+{\rm h.c.} ] -(1+|\Gamma|^2)\sum_j \hat S^z_j \hat S^z_{j+1}$, with $\Gamma\equiv \frac{1}{2}J_0(\Omega_{AB})e^{{\rm i}\Phi/2}$. Hence the perturbative corrections result in nearest neighbor interactions and staggered correlated hopping. 
Following similar arguments as those employed for the treatment of the spin-Peierls problem~\cite{GiamarchiBook}, one may show that the 
staggered correlated hopping becomes immediately relevant (in the renormalization group sense) for free hard-core particles, and hence any AB-dependent $\Gamma$ opens a (band insulator) gapped phase at half-filling for $U\to\infty$~(see Supplemental Material~\cite{SM} for details). 
A similar reasoning applies for higher half-integer fillings $\bar n +1/2$, by considering hard-core particles on top of a pseudo-vacuum with $\bar n$ particles per site.
Note that the Mott boundaries depend on $\Phi$ and hence varying $\Phi$ at constant $J/U$ results in gauge-induced 
phase transitions~(Fig.~\ref{fig:3}(b)), similar as the statistical transitions of Ref.~\cite{Keilmann2011}. In particular, 
for $\Phi\to\pi$ one observes a strong enhancement of the MI gaps. Half-integer and integer MI may be revealed by the appearance 
of density plateaus in the presence of a harmonic trap~\cite{Sherson2010}.

\emph{Vanishing on-site interaction.--} The effect of the density-dependent hopping is particularly relevant in the regime of vanishing interaction, $U/J\to 0$. 
In this regime, for the usual Hubbard model, the system becomes unstable for $\mu>-J$, i.e. any filling factor becomes possible~(note the bunching of 
curves of constant filling for $\Omega_{AB}=0$ in Fig.~\ref{fig:3}(c)). The presence of density-dependent hopping stabilizes the 
system at low fillings~(Fig.~\ref{fig:3}(c)). Moreover, the AB asymmetry results in a MI at half-filling even for $U/J=0$. 
This anomalous behavior results from the effective repulsive character of the gas even when $U=0$. 
This may be understood from the two-particle problem, which provides a useful description in the dilute limit~\cite{Kolezhuk11}. 
The effective scattering length~(in lattice spacing units) becomes of the form~\cite{SM}:
\begin{equation}
a(U\to0) = \frac{ \left[3+5 \cos(\Phi)\right]\left|\Gamma\right|^2 +2}{\left[5+3 \cos(\Phi)\right]\left|\Gamma\right|^2 -2},
\end{equation}
By comparison to a 1D Bose gas of particles with mass $m$ and contact-interaction one may extract an effective interaction strength $g=-2/(a m)$ ~\cite{Kolezhuk11}. The scattering length diverges for $|\Gamma|\to 1/2, \Phi\to0, 2\pi$ but remains finite and negative for any other phase $\Phi$ which coincides with the observation that the AB-correlated hopping Hubbard model behaves as a repulsively interacting system for small filling even in the limit of $U\to 0$. Incidentally, we would like to mention that this effect may be observed as well for the anyon model of Ref.~\cite{Keilmann2011}, although in that case the Mott plateau at half-filling is absent. 

\begin{figure} [t]
 \begin{center}
 \includegraphics[width =0.9\columnwidth]{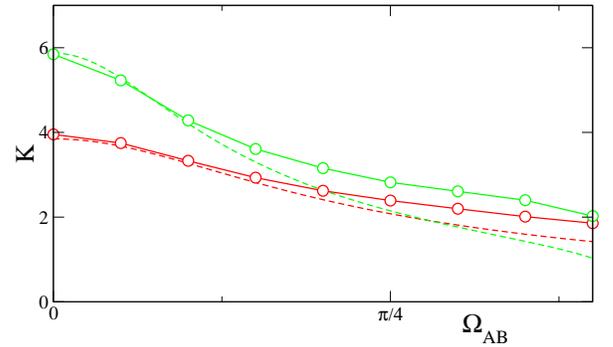} 
 \caption{ (Color online) Behavior of the Luttinger parameter $K$ as a function of $\Omega_{AB}$ for $U=J/2$ for $\rho_0=1.75$ (upper curves) and $\rho_0=0.75$ (lower curves). 
 Dashed lines indicate the analytical estimation~\eqref{eq:Analytic} in the weakly-interacting regime, whereas the circles denote our results obtained from DMRG calculations of the single-particle correlation function.}
 \label{fig:new}
 \end{center}
 \end{figure}

\emph{Correlation functions in the superfluid regime.--} The density-dependent gauge has important consequences for the correlations in the SF regime~\cite{SM}.
This is best understood by employing bosonization~\cite{GiamarchiBook}: $\hat b_j^\dag\rightarrow \sqrt{\rho(x_j)}e^{{-\rm i}(\theta(x_j)-\eta x_j)}$, with 
$\rho(x)=\rho_0-\frac{1}{\pi}\nabla\phi(x)+\rho_0\sum_{p\neq 0} e^{{\rm i}2p(\pi\rho_0+\phi(x))}$, $\rho_0$ the average density, and $x_j$ the position of site $j$. The fields $\theta(x)$ and $\phi(x)$ characterize the density and phase, respectively, 
whereas $\eta$ is for a global gaugeable phase shift. The bosonized Hamiltonian acquires the form~\cite{SM}:    
\begin{equation}
\hat H=\frac{u}{2\pi}\int dx \left [ K^{-1} (\partial_x\phi)^2+K (\partial_x\theta)^2+2\gamma (\partial_x\phi)(\partial_x\theta) \right ], 
\end{equation}
where $u$ is a velocity, $K$ is the Luttinger parameter, and $\gamma$ characterizes a mixing term that stems from the density-dependent Peierls phase. 
The decay of single particle correlations depends only on $K$ as $\langle \hat b_i^\dag b_j\rangle \propto |i-j|^{-1/2K}$~\cite{SM}. As depicted in Fig.~\ref{fig:new}, $K$ decreases with increasing $\Omega_{AB}$. This behavior can be understood already from the weak-coupling regime, in which $K$ may be determined analytically~\cite{SM}:
\begin{equation}
K^2=\frac{\pi^2\rho_0\tilde F(\rho_0)}{\frac{2U}{J}-{\cal R}\left ( \rho_0\frac{d^2\tilde F}{d\rho^2}(\rho_0) +2\frac{d\tilde F}{d\rho}(\rho_0) \right ) }
\label{eq:Analytic}
\end{equation} 
with ${\cal R}$ the real part, $\tilde F(\rho)=F(\rho)e^{-\mathrm{i}\arg(F(\rho_0))}$, and $F(\rho)=J_0(\Omega_{AB}\rho)e^{\mathrm{i}\Omega_{AB}\rho}$ for the AB model~(but the result may be generally 
applied to other forms of density-dependent tunneling, $F(\hat n_j)$).  Figure~\ref{fig:new} shows that our DMRG results are in excellent agreement with Eq.~\eqref{eq:Analytic}  
for small $\Omega_{AB}$, which corresponds to the weak-coupling limit.  The reduction of $K$ results on one hand from the trivial reduction of the 
hopping strength ($J\rightarrow J\tilde F(\rho_0)$), and on the other from a non-trivial contribution due to the density dependence~(denominator of $K$). 
The later stems from the effective repulsion discussed above. 
Note in particular, that a density-dependent Peierls phase, with $|F(\rho)|=1$, as that of Ref.~\cite{Keilmann2011}, would lead as well to 
strongly modified correlations characterized by a significant reduction of $K$~\cite{SM}. 
The modification of $K$ due to the density-dependent gauge may be directly probed by monitoring the form of the central momentum peak~\cite{Paredes2004}.

  \begin{figure} [t]
  \begin{center}
  \includegraphics[width =0.9\columnwidth]{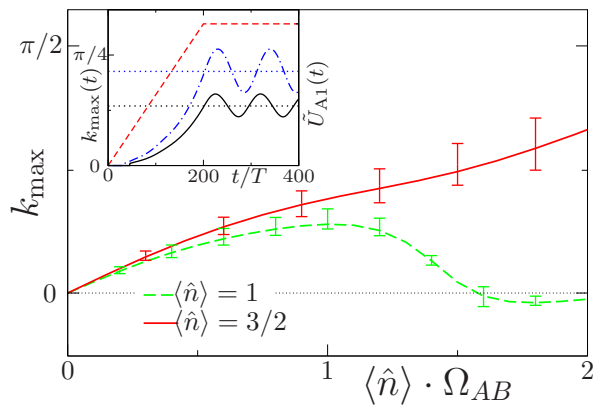} 
 \caption{(Color online) Quasi-momentum $k_{max}$ at which the quasi-momentum distribution of the B sublattice 
 is maximal as a function of $\Omega_{AB}\langle \hat n \rangle$ for $\omega=20J$ and $U=J$. 
 Solid~(dashed) lines denote the results obtained from the effective model~\eqref{eq:Heff} with $\langle \hat n \rangle =3/2$~($1$). 
 The error bars denote the uncertainty (time average and standard deviation for $200<t/T<400$) of $k_{max}(t)$ for the case of a linear ramp of 
 $\tilde U_{A1}$ with a ramp time of $\tau = 200 T$~(see text). 
 (inset) Solid and dash-dotted lines show $k_{max}(t)$ for $\langle\hat n\rangle=3/2$ with $\Omega_{AB}=0.4$ and $0.8$, whereas the dotted line indicates the 
 value of $k_{max}$ for the effective model~\eqref{eq:Heff}. We depict with a dashed line the ramp $\tilde{U}_{A1}(t)$.}
  \label{fig:5}
  \end{center}
  \end{figure}

\emph{Adiabatic preparation.--} We have focused above on the effective model~\eqref{eq:Heff}. 
As for shaken lattices~\cite{Poletti2011}, one may start from the ground-state without modulated interactions, and 
adiabatically increase $\tilde U_{A1}$. We have studied this preparation by means of time-evolving 
block decimation~(TEBD)~\cite{Daley2004} simulations of the dynamics of Eq.~\eqref{eq:HAB(t)} when applying 
a linear ramp $\tilde U_{A1}(t)=\frac{t}{\tau} \tilde U_{A1}$ for $t<\tau$, and constant afterwards \cite{SM}. 
Fig.~\ref{fig:5} depicts the value $k_{max}$ at which the momentum distribution is maximal, showing 
that the evolved momentum distribution is in very good agreement with that of the effective model.
Note that the drift $k_{max}$ is only linear with $\Omega_{AB}\langle \hat n \rangle$ for a sufficiently small value of $\Omega_{AB}\langle \hat n \rangle$. 
For larger $\Omega_{AB}\langle \hat n \rangle$ it presents a non-trivial density dependence, especially at low $\langle \hat n \rangle$, due to number fluctuations.

\emph{Detection.--} Whereas the density distribution of the effective model corresponds to that measured in the laboratory frame, the measurement of the momentum 
distribution in TOF presents some features that differ significantly from the shaken lattice case~\cite{Struck2012}.
First, since the lattice is not actually shaken, the overall momentum envelope resulting from the Fourier transform of the Wannier functions does not oscillate in time. 
Second, whereas the momentum distribution of the B sublattice measured in TOF corresponds to that of the effective model, the distribution of the A sublattice just coincides with that of the effective model (and also with that of the sublattice B) when $V(t)=0$.
For intermediate times, the phase appearing in the conversion between both reference frames 
leads to a broadening, and eventual blurring, of the TOF peaks \cite{SM}.

\emph{Outlook.--} Periodic interactions combined with Raman-assisted hopping
may create a density-dependent Peierls phase that results in non-trivial ground-state properties, characterized by  
a density-dependent momentum distribution, gauge-induced SF to MI transitions, the stabilization of the Hubbard model at vanishing interactions, and 
modified correlations in the SF phase. Although our discussion has focused on the specific case of the AB model, these peculiar properties are 
general for all models with a density-dependent Peierls phase~\cite{footnote-AB} (in the Supplemental Material~\cite{SM} we comment on the case of the anyonic model of Ref.~\cite{Keilmann2011}). 

The AB model may be extended to create a density-dependent gauge field in a square lattice, in which each row is an exact copy of the AB 
lattice as that discussed above, and rows are coupled by direct (not Raman-assisted) hops. Tilting the lattice, 
leads to a row-dependent $\langle \hat n \rangle$, and hence to a different Peierls phase at each row when 
modulating the interactions. In this way a finite flux may be produced in each plaquette, proportional 
to the density difference between neighboring rows. As a result, density dependent synthetic magnetic fields may be created, opening interesting possibilities that deserve further investigation.

\emph{Acknowledgements.--} We thank C. de Morais Smith, M. Di Liberto, M. Dalmonte, A. Eckardt, and U. Schneider for enlightening discussions. 
We acknowledge support by the cluster of excellence QUEST, the DFG Research Training Group 1729, and the 
SUTD start-up grant (SRG-EPD-2012-045).

\bibliographystyle{prsty}

\begin{thebibliography}{}

\bibitem{Dalibard2011} J. Dalibard, F. Gerbier, G. Juzeliunas, and P. \"Ohberg, Rev. Mod. Phys. {\bf 83}, 1523 (2011).
\bibitem{Goldman2013} N. Goldman, G. Juzeliunas, P. \"Ohberg abd I. B. Spielman, arXiv:1308.6533.
\bibitem{Lin2009} Y. J. Lin {\it et al.}, Phys. Rev. Lett. {\bf 102}, 130401 (2009).
\bibitem{Lin2009b} Y. J. Lin {\it et al.}, Nature {\bf 462}, 628 (2009).
\bibitem{Lin2011}  Y. J. Lin {\it et al.},  Nature Physics {\bf 7}, 531 (2011).
\bibitem{Lin2011b}  Y. J. Lin, K. Jim\'enez-Garc\'ia and I. B. Spielman,  Nature Physics {\bf 471}, 83 (2011).
\bibitem{Wang2012} P. Wang {\it et al.}, Phys. Rev. Lett. {\bf 109}, 095301 (2012).
\bibitem{Cheuk2012} L. W. Cheuk {\it et al.}, Phys. Rev. Lett {\bf 109}, 095302 (2012).
\bibitem{Zhang2012} J. Y. Zhang {\it et al.}, Phys. Lett. Lett. {\bf 109}, 115301 (2012).
\bibitem{Fu2013} Z. Fu {\it et al.} Nat. Phys. 10, 110 (2014).
\bibitem{Zhang2013} L. Zhang {\it et al.}, Phys. Rev. A {\bf 87}, 011601(R) (2013).
\bibitem{Qu2013} C. Qu, C. Hamner, M. Gong, C. Zhang and P. Engels, Phys. Rev. A {\bf 88}, 021604(R) (2013).
\bibitem{LeBlanc2013} L. J. LeBlanc {\it et al.}, New. J. Phys. {\bf 15}, 073011(2013).
\bibitem{Aidelsburger2011} M. Aidelsburger {\it et al.} Phys. Rev. Lett. {\bf 107}, 255301 (2011).
\bibitem{Jimenez-Garcia2012} K. Jim\'enez-Garc\'ia {\it et al.}, Phys. Rev. Lett. {\bf 108}, 225303 (2012).
\bibitem{Struck2012} J. Struck {\it et al.}, Phys. Rev. Lett. {\bf 108} 225304 (2012).
\bibitem{Aidelsburger2013} M. Aidelsburger {\it et al.}, Phys. Rev. Lett. {\bf 111},185301 (2013).
\bibitem{Miyake2013} H. Miyake, G.A. Siviloglou, C.J. Kennedy, W.C. Burton and W. Ketterle, Phys. Rev. Lett. {\bf 111},185302 (2013).
\bibitem{Levin2005} M. Levin and X. G. Wen, Rev. Mod. Phys. {\bf 77}, 871 (2005).
\bibitem{Kogut1983}  J. Kogut, Rev. Mod. Phys. {\bf 55}, 775 (1983).
\bibitem{Wiese2013}  U. J. Wiese, arXiv:1305.1602.
\bibitem{Cirac2010} J. I. Cirac, P. Maraner, and J. K. Pachos, Phys. Rev. Lett. {\bf 105}, 190403 (2010).
\bibitem{Zohar2011} E. Zohar and B. Reznik, Phys.Rev. Lett. {\bf 107}, 275301 (2011).
\bibitem{Kapit2011} E. Kapit and E. Mueller, Phys. Rev. A {\bf 83}, 033625 (2011).
\bibitem{Zohar2012} E. Zohar, J. I. Cirac, and B. Reznik, Phys. Rev. Lett. {\bf 109}, 125302 (2012).
\bibitem{Banerjee2012} D. Banerjee {\it et al.}, Phys. Rev. Lett. {\bf 109}, 175302 (2012).
\bibitem{Tagliacozzo2012} L. Tagliacozzo, A. Celi, P. Orland, and M. Lewenstein, Nat. Commun. 4, 2615 (2013). 
\bibitem{Zohar2013} E. Zohar, J. I. Cirac, and B. Reznik, Phys. Rev. Lett. {\bf 110}, 055302 (2013).
\bibitem{Banerjee2013} D. Banerjee {\it et al.}, Phys. Rev. Lett. {\bf 110}, 125303 (2013).
\bibitem{Zohar2013b} E. Zohar, J. I. Cirac, and B. Reznik, Phys. Rev. Lett. {\bf 110}, 125304 (2013).
\bibitem{Tagliacozzo2013} L. Tagliacozzo, A. Celi, P. Orland, M. W. Mitchell, and  M. Lewenstein, Nat. Commun. {\bf 4}, 2615 (2013).
\bibitem{Keilmann2011} T. Keilmann, S. Lanzmich, I. McCulloch, and M. Roncaglia, Nat. Commun. {\bf 2}, 361 (2011).
\bibitem{Edmonds2013} M.J. Edmonds, M. Valiente, G. Juzeliunas, L. Santos and P. Ohberg, Phys. Rev. Lett. {\bf 110}, 085301 (2013).
\bibitem{Eckardt2005} A. Eckardt, C. Weiss, and M. Holthaus, Phys. Rev. Lett. {\bf 95}, 260404 (2005).
\bibitem{Lignier2007} H. Lignier, \textit{et al.}, Phys. Rev. Lett. \textbf{99}, 220403 (2007).
\bibitem{Kierig2008} E. Kierig, U. Schnorrberger, A. Schietinger, J. Tomkovic and M.K. Oberthaler, Phys. Rev. Lett. {\bf 100}, 190405 (2008).
\bibitem{Zenesini2009} A. Zenesini, H. Lignier, D. Ciampini, O. Morsch and E. Arimondo, Phys. Rev. Lett. {\bf 102}, 100403 (2009).
\bibitem{Struck2011} J. Struck {\it et al.}, Science {\bf 333}, 996 (2011).
\bibitem{Chen2011} Y.-A. Chen {\it et al.}, Phys. Rev. Lett. {\bf 107}, 210405 (2011).
\bibitem{Ma2011} R. Ma {\it et al.}, Phys. Rev. Lett. {\bf 107}, 095301 (2011).
\bibitem{Gong2009} J. Gong, L. Morales-Molina and P. H\"anggi, Phys. Rev. Lett. {\bf 103}, 133002 (2009).
\bibitem{Rapp2012} \'A. Rapp, X. Deng, and L. Santos, Phys. Rev. Lett. {\bf 109}, 203005 (2012).
\bibitem{DiLiberto2013} M. Di Liberto, C. E. Creffield, G. I. Japaridze, and C. Morais Smith, Phys. Rev. A 89, 013624 (2014).

\bibitem{footnote-Jaksch} The laser arrangement is basically the same as that of 
[D. Jaksch and P. Zoller, New. J. Phys. {\bf 5}, 56 (2003)] proposed for the creation of a synthetic (static) magnetic field. 
However, here we do not demand a spatial dependence of the Rabi frequencies and the AB and BA lasers are switched on and off. The 
tilting must be sufficiently large to be resolved by the two different Raman pairs. 
The tilting must be also larger than the Raman-induced hopping rate and the interaction energy. Note also that the tilting should be chosen avoiding 
photon-assisted resonances~[C. Sias {\it et al.}, Phys. Rev. Lett. {\bf 100}, 040404 (2008)], which could result in a significant BA hopping even during the AB pulses. 

\bibitem{footnote-interaction} An even more intriguing phase space could emerge from assuming $U_{A0}\ne U_B$, however this goes beyond the scope of this current article. 

\bibitem{SM} See the Supplemental Material for details on periodically modulated interactions, the strongly-interacting limit, the calculation of the scattering length for vanishing interaction, 
time of flight imaging, correlation functions in the SF regime, and the numerical simulations.

\bibitem{footnote-anyon}The AB model resembles the anyon model of Ref.~\cite{Keilmann2011}, in which 
the inter-site hopping depends on the occupation of the left site. The model of Ref.~\cite{Keilmann2011} requires twice as many Raman lasers as the maximal 
occupation per site, and on-site interactions larger than the laser linewidth. The AB model works with
only one laser pair, and for interaction shifts smaller than the laser linewidth~(for $\langle \hat n \rangle=5$ and $U=0.2J$, the linewidth required must be larger than
$U\langle \hat n \rangle=J$; for $J$ of the order of tens of Hz this is a realistic assumption for typical linewidths~\cite{Miyake2013}). 
The AB model may be recast as an anyon model without Peierls phase by defining $\hat a_{2j}=e^{{\rm i}\Omega_{AB}\sum_{l<j} \hat n_{2l}} \hat b_{2j}$, and $\hat a_{2j+1}=e^{{\rm i}\Omega_{AB}\sum_{l\le j} \hat n_{2l}} \hat b_{2j+1}$, where the $\hat a$ operators fulfill, for $j'> j$, 
$e^{{\rm i}\Omega_{AB}}\hat a_{2j}^\dag \hat a_{2j'}=\hat a_{2j'}\hat a_{2j}^\dag$, 
$e^{{\rm i}\Omega_{AB}}\hat a_{2j}^\dag a_{2j'+1}=\hat a_{2j'+1}\hat a_{2j}^\dag$, $\hat a_{2j+1}^\dag \hat a_{2j'+1}=\hat a_{2j'+1}\hat a_{2j+1}^\dag$.

\bibitem{Schollwock2011} U. Schollw\"ock, Annals of Physics {\bf 326}, 96 (2011).

\bibitem{GiamarchiBook} Th. Giamarchi, {\it Quantum Physics in One Dimension}, (Oxford University Press, New York, 2004).

\bibitem{Sherson2010} J. Sherson {\it et al.}, Nature 467, {bf 68} (2010).

\bibitem{Kolezhuk11} A.~K.~Kolezhuk, F.~Heidrich-Meisner, S.~Greschner and T.~Vekua, Phys. Rev B {\bf 85}, 064420 (2012).

\bibitem{Paredes2004} B. Paredes {\it et al.}, Nature {\bf 429}, 277 (2004).

\bibitem{Poletti2011} D. Poletti and C. Kollath, Phys. Rev. A {\bf 84}, 013615 (2011).
\bibitem{Daley2004} A. J. Daley, C. Kollath, U. Schollw\"ock, and G. Vidal, J. Stat. Mech. (2004) P04005.

\bibitem{footnote-AB} Only the half-integer MI and the MI at vanishing interactions demand necessarily 
the AB asymmetry specific of the AB model.

\end{thebibliography}

\end{document}


\title{Supplementary material to \\ ``Density dependent synthetic gauge fields using periodically modulated interactions''}

\author{S. Greschner}
\affiliation{Institut f\"ur Theoretische Physik, Leibniz Universit\"at Hannover, Appelstr. 2, DE-30167 Hannover, Germany}

\author{G. Sun}
\affiliation{Institut f\"ur Theoretische Physik, Leibniz Universit\"at Hannover, Appelstr. 2, DE-30167 Hannover, Germany}

\author{D. Poletti}
\affiliation{Engineering Product Development, Singapore
University of Technology and Design, 20 Dover Drive, 138682 Singapore}

\author{L. Santos}
\affiliation{Institut f\"ur Theoretische Physik, Leibniz Universit\"at Hannover, Appelstr. 2, DE-30167 Hannover, Germany}
\maketitle

In this supplementary material, we provide additional details on the periodically modulated interactions, the strongly-interacting limit, 
the calculation of the scattering length of the AB-model, some aspects of the time of flight (TOF) imaging, 
correlation functions in the presence of density-dependent hopping, and the numerical simulation of real-time evolutions.

\renewcommand\thefigure{A\arabic{figure}}
\setcounter{figure}{0}

\section{A. Single-component Bose-Hubbard Hamiltonian with periodically modulated interactions}

We consider in this section single-component bosons in a 1D optical lattice with periodically modulated short-range interactions. 
Considering a large-enough gap between the first two Bloch bands, we may restrict the description of the system to a single band Bose-Hubbard Hamiltonian:
\begin{eqnarray}
 \hat H(t) = - J\sum_{\langle ij\rangle} \hat b_i^\dag \hat b_j  + \frac{U_0+U_1(t)}{2}\sum_i 
\hat n_i \left ( \hat n_i -1\right ),
  \label{eq:def:H(t)}
\end{eqnarray}
where $\hat b_i$ is the bosonic annihilation operator at site $i$, $\hat n_i =\hat b_i^\dag \hat b_i$, $J>0$ is the hopping rate, 
$\langle..\rangle$ denotes nearest neighbors, $U_1(t)=U_1(t+T)$, and $\int_t^{t+T} dt'\, U_1(t')=0$. 

We perform the transformation $|\psi'(t)\rangle=\hat R(t) |\psi(t)\rangle$, with 
$\hat R(t) = e^{ \mathrm{i}\frac{V(t)}{2} \sum_j \hat n_j \left ( \hat n_j -1\right ) }$, such that $\frac{d}{dt}V(t)=U_1(t)$~(note that $V(t)=V(t+T)$ since 
$U_1(t)$ is unbiased). In the transformed frame: $\mathrm{i}\hbar\partial_t|\psi'(t)\rangle=\hat H'(t)|\psi'(t)\rangle$, with 
$\hat H'= \hat R\hat H\hat R^\dag-\mathrm{i}\hbar\hat R\frac{d}{dt}\hat R^\dag$. 
Assuming a fast modulation, $\omega=2\pi/T\gg J/\hbar,U_0/\hbar$~\cite{footnote-heating}, we integrate the modulation to obtain the effective time-independent Hamiltonian
\begin{equation}
 \hat H_{\rm eff} = - \sum_{\langle ij\rangle} \hat b_i^\dag J_{\rm eff}(\hat n_i-\hat n_j) \hat b_j  + \frac{U_0}{2}\sum_i  \hat n_i \left ( \hat n_i -1\right ),
  \label{eq:def:Heff(t)}
\end{equation} 
with an effective density-dependent hopping $J_{\rm eff}(\Delta\hat n)=\frac{J}{T}\int_0^T dt \;e^{ \mathrm{i} V(t) \Delta\hat n }$ in the transformed frame~(see Refs.~\cite{Gong2009,Rapp2012} for further details).

We are interested in probing the effective model by TOF measurements. Note, however, that TOF measurements will monitor the evolution of the 
quasi-momentum distribution in the laboratory frame, 
$\rho_L(k,t)=\frac{1}{NL}\sum_{l,j} e^{-\mathrm{i}k(l-j)}\langle \psi(t)|\hat b_l^\dag \hat b_j |\psi(t)\rangle$, with $N$ the number of particles, and $L$ 
the number of sites. The single-particle correlation function in the laboratory frame fulfills: 
$\langle \psi(t)|\hat b_i^\dag \hat b_j |\psi(t)\rangle = \langle \psi'(t)|\hat b_i^\dag e^{\mathrm{i}V(t)(\hat n_i-\hat n_j)} \hat b_j |\psi'(t)\rangle$, 
and hence for general times 
$\rho_L(k,t)$ does not coincide with the quasi-momentum distribution of the effective model 
$\rho_E(k)=\frac{1}{NL}\sum_{l,j} e^{-\mathrm{i}k(l-j)}\langle \psi'|\hat b_l^\dag \hat b_j |\psi'\rangle$. 
We will be interested in stroboscopic measurements at times $t=nT$, with $n=0,1,\dots$, such that $\rho_{L}(k,nT)=\rho_E(k)$. 
This condition demands $V(0)=0$, fixing the gauge uncertainty~(we assume this 
gauge fixing henceforth). Hence measurements at times $t = n T$ allow to probe an effective model, $\hat{H}_{eff}$ (see Eq.(2) of the main text). In the following we consider for simplicity $U_1 (0) = 0$~.

Note that if the modulation starts at time $-T<-t_0<0$, $U_1(-t_0)=0$, then the time evolution between $t=-t_0$ and $t=0$ must 
be explicitly considered, i.e. the initial condition for the time 
evolution under the effective model is $|\psi'(0)\rangle={\cal T} e^{-{\rm i}\int_{-t_0}^0 \hat H'(t') dt'} |\psi(-t_0)\rangle$, where ${\cal T}$ denotes time ordering. 
The discrete time evolution at times $nT$ may be then evaluated, in a very good approximation for $\hbar\omega\gg U_0,J$, by evolving with $\hat{H}_{\rm eff}$ 
starting with the calculated $|\psi'(0)\rangle$. Note, however, that the measurements will probe a different effective model, due to the different~(shifted) 
form of $U_1(t)$, and hence of $V(t)$, and in turn of $J_{\rm eff}(\Delta\hat n)$.

For a sinusoidal modulation $U_1(t)=\tilde U_1 \sin(\omega t)$, $V(t)=\frac{\tilde U_1}{\omega}\left [ 1- \cos(\omega t) \right ]$, and hence 
$J_{\rm eff}(\Delta\hat n)=Je^{\mathrm{i}\Omega \Delta\hat n} J_0(\Omega \Delta\hat n)$, with $\Omega=\tilde U_1/\hbar\omega$ and $J_0$ the Bessel function of first kind. 
The stroboscopic measurement of $\rho_L(k,nT)$ allows hence to probe $\rho_E(k)$ for an effective model with 
a complex $J_{\rm eff}(\Delta\hat n)=|J_{\rm eff}(\Delta\hat n)|e^{\mathrm{i}\phi(\Delta\hat n)}$, with a quantum Peierls phase $\phi(\hat n_i-\hat n_j)=\Omega(\hat n_i-\hat n_j)$, 
dependent on the population difference between nearest sites. 

 \begin{figure}[t]
 \begin{center}
 \includegraphics[width =\columnwidth]{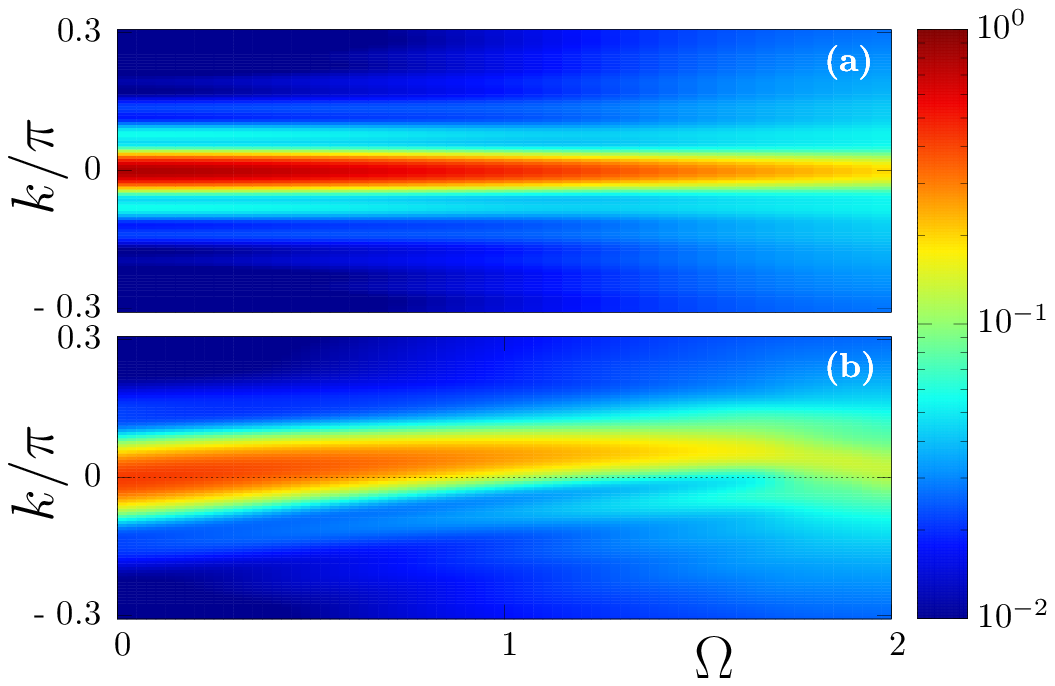}
 \caption{(a) Quasi-momentum distribution of the ground-state of~\eqref{eq:def:Heff(t)} with 
 $J_{\rm eff}(\Delta\hat n)=Je^{\rm{i}\Omega \Delta\hat n}J_0(\Omega \Delta\hat n)$ as a function of $\Omega$ 
 for $\langle \hat n \rangle=1$ and $U_0=J$. In order to exclude possible superfluid-to-insulator transitions
 we keep $J/J_0(\Omega)$ constant. (b) Same as (a) but with an on-site energy gradient, 
 $\epsilon \sum_j j n_j$ with $\epsilon=0.2 J$. DMRG calculations~\cite{Schollwock2011} were performed in $L=36$ sites, with 
 a maximal site occupation of $6$ bosons.}
 \label{fig:SBH_phase}
 \end{center}
 \end{figure}

The hopping may hence acquire a density-dependent Peierls phase but it may be gauged out by defining 
new bosonic operators $\hat B_j\equiv e^{-{\rm i}\Omega\hat n_j}\hat b_j$. Hence, the complex hopping does not affect the ground-state phase diagram~\cite{footnote-caveat}.
Moreover, the appearance of this phase does not result in an overall shift of $\rho_E(k)$. This may be understood by realizing that by construction $J_{\rm eff}(-\Delta\hat n)=J_{\rm eff}(\Delta\hat n)^*$, and hence $\phi(-\Delta\hat n)=-\phi(\Delta\hat n)$. For an homogeneous superfluid, the number difference between neighboring sites presents quantum fluctuations around a zero mean, and 
the quantum Peierls phases acquire a stochastic character, variating randomly from bond to bond between positive and negative values. 
As a result the system experiences an effective decoherence. We illustrate this effect in Fig.~\ref{fig:SBH_phase}(a), where we show DMRG results for $\rho_E(k)$. Note that when increasing $\Omega$, $\rho_E(k)$ broadens, and may even saturate the Brillouin zone, as a consequence of the dephasing.

A net drift of the quasi-momentum distribution may be however achieved in the presence of a density gradient, which may in turn result from a lattice tilting. 
Although the created phase still depends on population differences, the density gradient 
$\langle \hat n_j - \hat n_{j+1} \rangle  \neq 0$ leads to a non-zero average Peierls phase. This is illustrated in Fig.~\ref{fig:SBH_phase}(b), where 
we show that a density gradient results in a net drift of the momentum distribution, in addition to the broadening mentioned above.

\section{B. Creation of arbitrary density dependent Peierls phases}

In the main text the derivation of the AB-model is described for the case $U_{A1}(0<t<T/2) = - U_{A1}(T/2<t<T) = \tilde{U}_{A1} \sin (\omega_{AB} t)$ which leads to the density dependent hopping amplitude and phase $\tilde{J}_{AB}(\hat{n}_{2j}) = \frac{J}{2} J_0(\Omega_{AB} \hat{n}_{2j}) e^{\mathrm{i} \Omega_{AB} \hat{n}_{2j}} = \tilde{J}_{BA}(\hat{n}_{2j})^*$. So here the phase is always strictly coupled to the modulus of the hopping.

One may choose more generally $U_{A1}(0<t<T/2) = \tilde{U}_{A1} \sin(\omega_{AB} t + \phi_1)$ and $U_{A1}(T/2<t<T) = \tilde{U}_{A1} \sin(\omega_{AB} t + \phi_2)$. Note that $\phi_1=0$, $\phi_2=\pi$ reproduces the case shown in figure 1 of the main text. 
The effective tunneling is given by $\tilde{J}_{AB}(\hat{n}_{2j}) = \frac{J}{2} J_0(\Omega_{AB} \hat{n}_{2j}) e^{\mathrm{i} \Omega_{AB} \cos (\phi_1) \hat{n}_{2j}}$ and  $\tilde{J}_{BA}(\hat{n}_{2j}) = \frac{J}{2} J_0(\Omega_{AB} \hat{n}_{2j}) e^{\mathrm{i} \Omega_{AB} \cos(\phi_2) \hat{n}_{2j}}$. 
A unitary gauge transformation $b_{2j}^\dagger \to b_{2j}^\dagger e^{-\mathrm{i} (\Phi_{AB}+\Phi_{BA})/2 \hat{n}_{2j}}$ may be used to obtain 
$\tilde{J}_{AB}(\hat{n}_{2j}) = \frac{J}{2} J_0(\Omega_{AB} \hat{n}_{2j}) e^{\mathrm{i} \Phi/2 \hat{n}_{2j}} = \tilde{J}_{BA}(\hat{n}_{2j})^*$ in Eq.~(4) of the main text.
Hence, $\Phi = \Phi_{AB}-\Phi_{BA} = \Omega_{AB} [\cos(\phi_1) - \cos(\phi_2)]$ may be changed keeping the hopping modulus unaffected as in Fig.~3(b) of the main text.

\section{C. Strongly interacting limit}

In the following we discuss briefly the physics in the limit of strong interactions, $U\to\infty$. As described in the main text, in this limit one can reduce the description to the manifold of $0$ and $1$ particles per site and introduce an effective spin-$\frac{1}{2}$ Hamiltonian $H_{1/2}$ in perturbation theory up to second order $J/U$ as given in the main text. This Hamiltonian may be rewritten as
\begin{align*}
\hat H_{1/2} &= -J \sum_j [\hat S^+_j\hat S^-_{j+1}+{\rm h.c.}] \\
&+ J_c \sum_j [\hat S^+_{j}  \left(\frac{1}{2} + \hat S^z_{j+1}\right) \hat S^-_{j+2} +{\rm h.c.}] +\\
&+ J_s \sum_j (-1)^j [\hat S^+_{j} \left(\frac{1}{2} + \hat S^z_{j+1}\right) \hat S^-_{j+2} +{\rm h.c.}] +\\
&+ \Delta \sum_j \hat S^z_j \hat S^z_{j+1}
\end{align*}
with coefficients $J_c=\frac{J^2}{U}\frac{1+\Gamma^2}{2}$, $J_s=\frac{J^2}{U}\frac{1-\Gamma^2}{2}$ and $\Delta=-\frac{J^2}{U}(1+|\Gamma|^2)$. Using the standard bosonization dictionary \cite{GiamarchiBook} the continuum limit of this Hamiltonian may be expressed as a sine-Gordon Hamiltonian of the density and phase fluctuations $\theta(x)$ and $\phi(x)$. It is precisely the staggered (next-nearest-neighbor) hopping that introduces at half filling a spin-Peierls like term $\sim \sin{2 \phi(x)}$ which becomes relevant for Luttinger-liquid parameters $K<2$. That is why at half filling we observe the immediate opening of band insulator gap for arbitrarily small tunneling $J/U$ which is consistent with our numerical simulations. The $S^z-S^z$-interaction contributes with $\sim \cos{4 \phi(x)}$ terms, which are irrelevant for $K>1/2$.

The opening of a gap may be also understood in an easier way if we just consider the correlated hopping parts $S^+_{j} \hat S^z_{j+1}\hat S^-_{j+2}$ of the second order perturbation, since this part may be analytically solved by mapping to free fermions:
\begin{align*}
\hat H_{1/2}^{\rm sf}&= -J \sum_j \hat c^\dag_j\hat c_{j+1} + J_c \sum_j \hat c^\dag_{j} \hat c_{j+2} + \\
&+ J_s \sum_j (-1)^j \hat c^\dag_{j} \hat c_{j+2} +{\rm h.c.}
\end{align*}
Here one finds the spin-Peierls like band-gap opening $\sim |J_s|$ at half filling.

\section{D. The two particle scattering problem}

In the following we provide a detailed description of the calculation of the two-particle scattering length for the AB-model as given in Eq.(3) of the main text. A general bosonic two particle state is given by
\[
\left | \Psi_Q \right \rangle =  \left [ \sum_x \frac{c_{x,x}}{ \sqrt{2}} \left(b_x^\dagger\right)^2 + \sum_{x,y>x} c_{x,y} b_x^\dagger b_{y}^\dagger\right ] \left|0\right\rangle,
\]
where $|0\rangle$ is the vacuum. Due to the conservation of total momentum in the scattering process one can express the amplitudes as $c_{x,x+r} = C_r \mathrm{e}^{{\rm i} Q (x+\frac{r}{2})}$ for $x$ in one of the $A$ sites and $c_{x,x+r} = D_r \mathrm{e}^{{\rm i} Q (x+\frac{r}{2})}$ for $x\in B$. Here $Q=q_1+q_2$, the total momentum~(below we employ $q=(q_1-q_2)/2$ as the half relative momentum).  The Schr\"odinger equation $\hat{H}^{AB}_{\rm eff}\left|\Psi\right>= \epsilon \left|\Psi\right>$ for the two particle problem leads to the following system of coupled equations for the amplitudes $C_r$ and $D_r$ with $\Gamma\equiv \frac{1}{2}J_0(\Omega_{AB})e^{\mathrm{i}\Phi/2}$
\begin{widetext}
\begin{align}
(\epsilon - U) C_0 &= - \sqrt{2} J \left|\Gamma\right| \left(D_1\, e^{\mathrm{i}(Q-\Phi)/2} + C_1\, e^{-\mathrm{i} (Q-\Phi)/2}\right) \nonumber\\
(\epsilon - U) D_0 &= - \sqrt{2} J \left|\Gamma\right| \left(C_1\, e^{\mathrm{i} Q/2} + D_1\, e^{-\mathrm{i} Q/2 }\right) \nonumber\\
\epsilon C_1 &= - \sqrt{2} J \left|\Gamma\right| \left( C_0\, e^{\mathrm{i}(Q-\Phi)/2} + D_0\, e^{-\mathrm{i} Q/2 } \right) -J/2\left( C_2\, e^{-\mathrm{i} Q/2 } + D_2\, e^{\mathrm{i} Q/2}\right)\nonumber\\
\epsilon D_1 &= - \sqrt{2} J \left|\Gamma\right|\left( C_0\, e^{-\mathrm{i}(Q-\Phi)/2} + D_0\, e^{\mathrm{i} Q/2} \right) -J/2 \left( C_2\, e^{\mathrm{i} Q/2 } + D_2\, e^{-\mathrm{i} Q/2}\right)\nonumber\\
\epsilon C_{r\ge 2} &= -J/2 \left( C_{r-1}\, e^{\mathrm{i} Q/2} + C_{r+1}\, e^{-\mathrm{i} Q/2} + D_{r-1}\, e^{-\mathrm{i} Q/2} + D_{r+1}\, e^{\mathrm{i} Q/2} \right) \nonumber\\
\epsilon D_{r\ge 2} &= -J/2 \left( D_{r-1}\, e^{\mathrm{i} Q/2} + D_{r+1}\, e^{-\mathrm{i} Q/2} + C_{r-1}\, e^{-\mathrm{i} Q/2} + C_{r+1}\, e^{\mathrm{i} Q/2} \right) \nonumber
\end{align}
\end{widetext}
The energy of the two scattered particles is given by $\epsilon = -2 J \cos (q) \cos (Q/2)$. In order to extract scattering properties we solve this set of equations with the ansatz $C_r = e^{-\mathrm{i} q r} + v e^{\mathrm{i} q r} + \beta \alpha^r$ and $D_r = e^{-\mathrm{i} q r} + v e^{\mathrm{i} q r} - \beta \alpha^r$ for $r>1$. The equations for $r>2$ can be solved by this ansatz if $2 \mathrm{i} \alpha \cos (q) \cos (Q/2) = (-1 + \alpha^2) \sin (Q/2)$. We choose $|\alpha|<1$ and solve the remaining four equations for $C_0, D_0, v \text{ and } \beta$. Since the $\alpha$ part decays exponentially fast, we can extract the scattering length $a = -\lim_{q\to0} \partial_q \delta$ with $v = \mathrm{e}^{2 {\rm i} \delta}$ which after some algebra results in Eq.(3) of the main text.

\section{E. Time of flight imaging}
 
\begin{figure} [t]
 \begin{center}
 \includegraphics[width =\columnwidth]{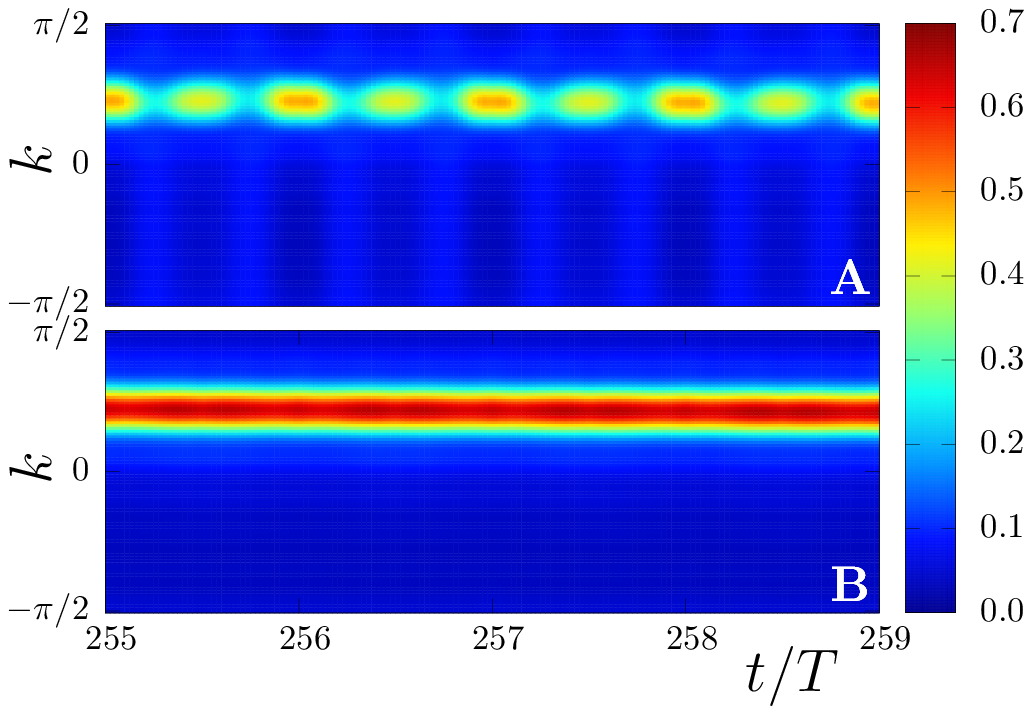} 
 \caption{Quasi-momentum distribution of the $A$ and the $B$ components in the laboratory frame as a function of time for $\langle \hat n\rangle=3/2$ 
 and $\Omega_{AB}=0.8$, and same parameters as those of Fig.5 of the main text.}
 \label{fig:SAB_mom_ramp}
 \end{center}
 \end{figure}

As in recent experiments on shaken lattices~\cite{Struck2012}, the shifted quasi-momentum distribution $\rho_E(k)$, 
may be detected in TOF experiments. However, as mentioned in the main text, the relation between the quasi-momentum distribution of the 
effective model and TOF imaging presents some features that differ significantly from the shaken lattice case. Interestingly, since atoms at sites $A$ and $B$ belong to different species, it is actually possible to visualize the quasi-momentum distribution of atoms in state $|1\rangle$ and $|2\rangle$ separately~(see Fig.~\ref{fig:SAB_mom_ramp}). 
Note that for the B sublattice, $\langle \psi (t)| \hat b_{2i+1}^\dag \hat b_{2j+1} |\psi(t)\rangle=\langle \psi'| \hat b_{2i+1}^\dag \hat b_{2j+1} |\psi'\rangle$, and hence the quasi-momentum distribution observed in TOF will be exactly the same as that of the effective model at any time. In contrast, for the A sublattice 
$\langle \psi (t)| \hat b_{2i}^\dag \hat b_{2j} |\psi(t)\rangle=\langle \psi'| \hat b_{2i}^\dag e^{\mathrm{i}V(t)(\hat n_{2i}-\hat n_{2j})/\hbar} \hat b_{2j} |\psi'\rangle$. 
As a result, the quasi-momentum distribution of the $A$ sublattice just coincides with that of the effective model~(and also with that of the sublattice B) at times $t=nT$. 
For intermediate times, the phase appearing in the conversion between both reference frames 
leads to a broadening, and eventual blurring, of the TOF peaks~(Fig.~\ref{fig:SAB_mom_ramp}). Note that this blurring is in itself a result of the number-dependence 
of the effective model, being related with the stochastic phase discussed in Sec. A.

\section{F. Correlation functions in the superfluid regime}

In this section we provide additional details concerning the bosonization of the AB model, and the calculation of correlation functions in the SF regime.
We consider the Hamiltonian:
\begin{eqnarray}
\hat H_T&=&-\frac{J}{2} \sum_j \left\{\hat b_{2j}^\dag F[\hat n_{2j}] \hat b_{2j+1}+\hat b_{2j}^\dag F[\hat n_{2j}]^* \hat b_{2j-1}+H.c.\right \} \nonumber \\
&+&\frac{U_0}{2}\sum_j \hat n_j(\hat n_1-1), 
\end{eqnarray}
which becomes the AB model in the main text for $F(x)= {\cal J}_0(\Omega_{AB} x)e^{-\mathrm{i}\Omega_{AB} x}$.  
We employ the bosonization: $b_j^\dag \rightarrow \rho(x_j)^{1/2} e^{-\mathrm{i}(\theta(x_j)-\eta x_j)}$, with $\rho(x)\simeq \rho_0-\frac{1}{\pi}\partial_x\phi$ (neglecting
the contribution of higher harmonics), $\rho_0$ the average density, and $x_j$ the position of site $j$. Note the displacement $\eta=\arg[F(\rho_0)]$ that results from the presence of a complex hopping. 
This displacement is introduced to remove linear terms in $\partial_x\theta$,  and leads to an overall drift of the momentum distribution. 
The Hamiltonian~(up to irrelevant constants and terms proportional to $\partial_x \phi$ that may be reabsorbed in the chemical potential) 
becomes of the form:
\begin{equation}
\hat H=\frac{u}{2\pi}\int dx \left [ K^{-1} (\partial_x\phi)^2+K (\partial_x\theta)^2+2\gamma (\partial_x\phi)(\partial_x\theta) \right ], 
\label{eq:HB}
\end{equation}
where in the weak-coupling regime: 
\begin{eqnarray}
\frac{uK}{2\pi}&=&\frac{J\rho_0}{2}  \tilde F(\rho_0), \\
\frac{u}{2\pi K}&=&\frac{U_0}{2\pi^2}-\frac{J}{2\pi^2}{\cal R} \left [\rho_0\frac{d^2\tilde F(\rho_0)}{d\rho^2}+2\frac{d\tilde F(\rho_0)}{d\rho}\right ], \\
\frac{u\gamma}{2\pi}&=&-\frac{J\rho_0}{2\pi}{\cal I} \left [ \frac{d\tilde F(\rho_0)}{d\rho}\right ],
\end{eqnarray}
with ${\cal R}$~(${\cal I}$)  the real~(imaginary) part. Note that $\gamma\neq 0$ only if the hopping is density-dependent and complex, i.e. 
in the presence of a density-dependent gauge field. 

In the strong-coupling regime the particular relation between the microscopic parameters and 
the coefficients of the low-energy Hamiltonian may be modified, but the form of the bosonized Hamiltonian~\eqref{eq:HB} is preserved.
We may hence evaluate correlation functions using the standard formalism, see e.g Ref.~\cite{GiamarchiBook}.

In particular, $\langle (\phi(x,\tau)-\phi(0,0))^2\rangle=K F_1(x,\tau)$ (with $\tau$ the imaginary time) and $\langle (\theta(x,\tau)-\theta(0,0))^2\rangle=K^{-1} F_1(x,\tau)$, 
where (introducing a cut-off length $\chi$ that may be equated to the lattice spacing): 
\begin{equation}
F_1(x,\tau)=\frac{1}{2\pi}\int_{-\infty}^{\infty} dk e^{-\chi |k|} \int_{-\infty}^{\infty} d\omega \frac{1-\cos(kx-\omega u\tau)}{k^2+(\omega-\mathrm{i}\gamma k)^2}.
\end{equation}
Note that compared to the expression with $\gamma=0$~\cite{GiamarchiBook}, the only effect of the mixing term $(\partial_x\phi)(\partial_x\theta)$ 
consists in a frequency shift $\omega\rightarrow \omega-\mathrm{i}\gamma k$. 
For $|\gamma|<1$~\cite{footnote-gamma}
\begin{widetext}
\begin{equation}
F_1(x,\tau)=\frac{1}{2}\ln \left [ \frac{\left (x+\mathrm{i}(\chi + u\tau(1+|\gamma|))\right )\left ( x-\mathrm{i}(\chi + u\tau(1-|\gamma|)) \right )}{\chi^2} \right ]
\end{equation}
\end{widetext}

For $\tau\rightarrow 0^+$, and $x\gg\chi$, we obtain $F_1(x)=\ln |x|$, i.e. the dependence found for $\gamma=0$\cite{GiamarchiBook}. 
As a consequence the single particle correlation acquires the standard ($\gamma$-independent) form:
$\hat b_i^\dag \hat b_j \sim |i-j|^{1/2K}$. Similarly, one may evaluate the density-density correlation which acquires as well the standard form:
$\langle \hat n_i \hat n_j \rangle =\rho_0^2-\frac{K}{2\pi^2x^2}$ (neglecting oscillatory terms).

\begin{figure} [t]
\begin{center}
\includegraphics[width =0.8\columnwidth]{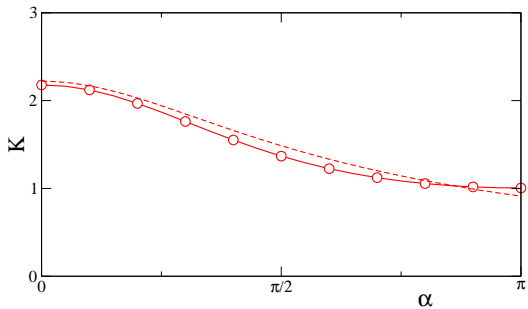} 
\caption{Dependence of the Luttinger parameter $K$ as a function of $\alpha$ for the anyon model of Ref.~\cite{Keilmann2011}, for $\rho_0=0.25$ and $U_0=t$. 
The circles denote our numerical results obtained from DMRG calculations of the single-particle correlation function, 
whereas the dashed line depicts the analytical curve~\eqref{eq:Anyon}.}
\label{fig:A3}
\end{center}
\end{figure}

Note that the arguments above are not specific for the AB model. Any density-dependent hopping would result in a Hamiltonian of the form~\eqref{eq:HB}. 
In particular for the anyon Hubbard model discussed in Ref.~\cite{Keilmann2011}: 
$\hat H=-t\sum_j \left\{ \hat b_j^\dag e^{\mathrm{i}\alpha \hat n_j} \hat b_{j+1}+H.c. \right \}+\frac{U_0}{2}\sum_j \hat n_1 (\hat n_j-1)$, 
one obtains in the weakly-interacting regime:
\begin{equation}
K^2=\frac{\pi^2}{\alpha^2+\frac{U_0}{2\rho_0 t}}, 
\label{eq:Anyon}
\end{equation}
which, as shown in Fig.~\ref{fig:A3} matches very well with our numerical results for $\rho_0=0.25$ and $U_0=t$.
Note that in this case a density-dependent Peierls phase, without any associated decrease of the hopping strength, results as well 
in a non-trivial decay of $K$.

\section{G. Details of the numerical simulation of real time evolutions}

For the dynamical calculations of Fig.~5 of the main text, and Fig.~\ref{fig:SAB_mom_ramp} we have used TEBD\cite{Vidal03} calculations for $16$ sites with up to $300$ states, and a maximal site occupation of $4$ bosons. As in Ref.~\cite{Poletti2011}, we may simulate rather long evolution times~($t\sim 400 T$) due to the 
quasi-adiabatic character of the dynamics. We have carried out our TEBD simulations for time steps $dt=T/400$ and
$m=300$ matrix states, which compare well to simulations with $dt=T/600$ and
$m=400$, showing the convergence of the results. Smaller system sizes, with a correspondingly decreased ramping and
evolution time, display very similar behavior and error-bars. The non-adiabaticity of the finite ramping time leads to oscillations in the expectation value of $k_{max}$ after the ramping procedure. The time-average and standard deviation are shown as points and error-bars in Fig.~5 of the main text and compare very well to the ground-state expectation.

\bibliographystyle{prsty}